# Affine Coxeter group $W_a(A_4)$, Quaternions, and Decagonal Quasicrystals

Mehmet Koca[a)], Nazife O. Koca[b)]
Department of Physics, College of Science, Sultan Qaboos University
P.O. Box 36, Al-Khoud, 123 Muscat, Sultanate of Oman
and
Ramazan Koc[c)]
Department of Physics, Gaziantep University, 27310, Gaziantep, Turkey

**Abstract**

We introduce a technique of projection onto the Coxeter plane of an arbitrary higher dimensional lattice described by the affine Coxeter group. The Coxeter plane is determined by the simple roots of the Coxeter graph $I_2(h)$ where $h$ is the Coxeter number of the Coxeter group $W(G)$ which embeds the dihedral group $D_h$ of order $2h$ as a maximal subgroup. As a simple application we demonstrate projections of the root and weight lattices of $A_4$ onto the Coxeter plane using the strip (canonical) projection method. We show that the crystal spaces of the affine $W_a(A_4)$ can be decomposed into two orthogonal spaces whose point groups is the dihedral group $D_5$ which acts in both spaces faithfully. The strip projections of the root and weight lattices can be taken as models for the decagonal quasicrystals. The paper also revises the quaternionic descriptions of the root and weight lattices, described by the affine Coxeter group $W_a(A_3)$, which correspond to the face centered cubic (fcc) lattice and body centered cubic (bcc) lattice respectively. Extensions of these lattices to higher dimensions lead to the root and weight lattices of the group $W_a(A_n)$, $n \geq 4$. We also note that the projection of the Voronoi cell of the root lattice of $W_a(A_4)$ describes a framework of nested decagram growing with the power of the golden ratio recently discovered in the Islamic arts.

[a)]electronic-mail: kocam@squ.edu.om
[b)]electronic-mail: nazife@squ.edu.om
[c)]electronic-mail: koc@gantep.edu.tr

## 1. Introduction

After the seminal paper of D. Shechtman et.al.[1] on the discovery of the long- range order of the rapidly solidified Al-Mn alloy exhibiting an icosahedral point symmetry, the phenomenon has been defined as a phase of quasicrystallographic structure of matter [2-3]. After the first discovery there has been an increasing interest in the quasicrystallography. Recent developments show that one can compose metallic alloys which display 5-fold, 8-fold, 10-fold, 12-fold, 18-fold symmetries. For a general exposition see the references on quasicrystallography [4]. Decagonal quasicrystals have been extensively studied in the references [5-6].

A complete mathematical description of the quasicrystals, contrary to the crystalline matter which has translational invariance, is still lacking. However, there has been a considerable progress by employing geometrical methods such as Penrose tiling [7] for the study of quasicrystallography with fivefold and tenfold symmetries or the algebraic description of the Penrose tiling by using pentagrid technique suggested by N. G. de Brujin [8]. By the orthogonal projection technique of the higher dimensional lattices one can describe the quasicrystalline structures in lower dimensions [9]. It was already noted by Shcherbak [10] that the Coxeter-Weyl groups $W(E_8)$, $W(D_6)$, and $W(A_4)$ possess the noncrystallographic Coxeter groups $W(H_4)$, $W(H_3)$, and $W(H_2)$ as maximal subgroups respectively. The orthogonal projection technique of the relevant root lattices, namely, $W(D_6) \rightarrow W(H_3)$ and $W(A_4) \rightarrow W(H_2)$ has been suggested [11] for the description of the 3D quasicrystallography with icosahedral point symmetry and the decagonal quasicrystallography in 2D respectively. Later the technique has also been applied to the $A_4$ weight lattice [12].

The long-range symmetric order with local decagonal symmetry shows itself also in Islamic decorative arts. Lu and Steinhardt [13] have shown that some medieval Islamic decorations can be explained in terms of tiling with five prototiles preserving local decagonal symmetry. Another very interesting work on the Islamic decorations indicates that there exists an underlying decagonal cartwheel structure as advocated by Al Ajlouni [14].

In this work we study the orthogonal projection of the root and weight lattices of $A_4$ with a different technique and show that it has a close correspondence with the pentagrid method of de Brujin [8] and the orthogonal projection of the Voronoi cell of the root lattice exhibits the nested cartwheel structure as suggested by Al Ajlouni [14]. We note that the projection of the hypercube in D5 can be explained in terms of the projection of the $B_5$ lattice generated by its short roots with a Coxeter-Weyl point group $W(B_5)$ which admits the group $W(A_4)$ as a maximal subgroup. We also note that $W(A_4)$ polytopes and their dual polytopes [15] play a crucial role in the study of decagonal quasicrystals. Moreover its description in terms of icosians indicates the richness of the $A_4$ lattices. We point out the importance of the $A_n$ ($n=1, 2, 3, 4, 5,...$) lattices by first studying the $A_3$ lattices. For a general study of the lattices and their symmetries we recommend the book by Conway and Sloane [16].

We organize the paper as follows. In Section 2 we introduce quaternionic description of the Coxeter-Dynkin diagram $A_3$ representing the tetrahedral symmetry which can be extended to the octahedral symmetry by the Dynkin diagram symmetry [17]. We point out that the Voronoi cell (Wigner-Seitz cell) is the rhombic dodecahedron of the root lattice equivalent to the face centered cubic lattice (fcc). Similarly, the Voronoi cell (truncated octahedron) is the unit cell of the weight lattice $A_3^*$ known as the body centered cubic lattice (bcc). Both lattices possess the point octahedral symmetries. In Section 3 we introduce a general technique as for the projection of a higher dimensional lattice described by the affine Coxeter group onto the Coxeter plane. We construct the Coxeter –Weyl group $W(A_4)$ in terms of quaternions which was well studied in reference [18] as a subgroup of the Coxeter group $W(H_4)$. For a further study of the quaternionic representation of the root lattice of $A_4$ and similar lattices we refer the reader to the reference [19]. The polytopes and dual polytopes [20] of $W(A_4)$ play an important role in the projection process which involves the Voronoi cells of the root and weight lattices. Section 4 is devoted to the study of the orthogonal projection of the lattices onto the Coxeter plane defined by the root system of the dihedral group generated by the Coxeter diagram $H_2$. Employing the strip projection technique we illustrate the point distributions of the projected lattice points manifestly displaying the dihedral symmetry $D_5$. Finally in Section 5 we present a discussion on the resemblance of the projected copy of the Voronoi cell of the root lattice $A_4$ with the basic structure of the Islamic design leading to many decorative patterns with dihedral symmetry $D_5$.

## 2. Construction of the affine Coxeter group $W_a(A_3)$ in terms of quaternions

Let $q = q_0 + q_i e_i$, $(i = 1, 2, 3)$ be a real unit quaternion with its conjugate defined by $\bar{q} = q_0 - q_i e_i$ and the norm $q\bar{q} = \bar{q}q = 1$. The quaternionic imaginary units satisfy the relations

$$e_i e_j = -\delta_{ij} + \varepsilon_{ijk} e_k, \quad (i, j, k = 1, 2, 3) \tag{1}$$

where $\delta_{ij}$ and $\varepsilon_{ijk}$ are the Kronecker and Levi-Civita symbols and summation over the repeated indices is implicit. They form a group isomorphic to the unitary group $SU(2)$. With the definition of the scalar product

$$(p, q) = \frac{1}{2}(\bar{p}q + \bar{q}p) = \frac{1}{2}(p\bar{q} + q\bar{p}), \tag{2}$$

quaternions generate the four-dimensional Euclidean space.

The Coxeter diagram $A_3$ with its quaternionic roots [17] is shown in Figure 1.

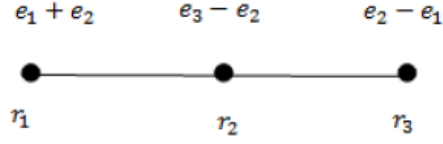

**Fig.1.** Coxeter-Dynkin diagram $A_3$ with quaternionic simple roots

The Cartan matrix of the Coxeter diagram $A_3$ and its inverse matrix (Gram matrix for the reciprocal lattice) are respectively given by

$$C = \begin{bmatrix} 2 & -1 & 0 \\ -1 & 2 & -1 \\ 0 & -1 & 2 \end{bmatrix}, \quad C^{-1} = \frac{1}{4}\begin{bmatrix} 3 & 2 & 1 \\ 2 & 4 & 2 \\ 1 & 2 & 3 \end{bmatrix}. \quad (3)$$

The simple roots $\alpha_i$ and their duals $\omega_i$ satisfy the scalar product relations

$$(\alpha_i, \alpha_j) = C_{ij}, \quad (\omega_i, \omega_j) = (C^{-1})_{ij}, \quad \text{and } (\alpha_i, \omega_j) = \delta_{ij}, \quad (i, j = 1, 2, 3). \quad (4)$$

They can be written as linear combinations of each other:

$$\alpha_i = C_{ij}\omega_j, \quad \omega_i = (C^{-1})_{ij}\alpha_j. \quad (5)$$

Let $\alpha$ be an arbitrary root given in terms of quaternions. Then the reflection of an arbitrary vector $\lambda$ with respect to the plane orthogonal to the root $\alpha$ is given by [21]

$$r\lambda = -\frac{\alpha}{\sqrt{2}}\bar{\lambda}\frac{\alpha}{\sqrt{2}} \equiv [\frac{\alpha}{\sqrt{2}}, -\frac{\alpha}{\sqrt{2}}]^*\lambda. \quad (6)$$

Therefore the generators of the Coxeter-Weyl group $W(A_3) \approx T_d \approx S_4$ are given by

$$r_1 = [\frac{1}{\sqrt{2}}(e_1 + e_2), -\frac{1}{\sqrt{2}}(e_1 + e_2)]^*,$$
$$r_2 = [\frac{1}{\sqrt{2}}(e_3 - e_2), -\frac{1}{\sqrt{2}}(e_3 - e_2)]^*, \quad (7)$$
$$r_3 = [\frac{1}{\sqrt{2}}(e_2 - e_1), -\frac{1}{\sqrt{2}}(e_2 - e_1)]^*.$$

The group elements of the Coxeter group which is isomorphic to the tetrahedral group of order 24 can be written compactly by the set

$$W(A_3) \approx T_d \approx S_4 = \{[p,\bar{p}] \oplus [t,\bar{t}]^*\}, \ p \in T, \ t \in T'. \tag{8}$$

Here $T$ and $T'$ represent respectively the sets of quaternions

$$T = \{\pm 1, \pm e_1, \pm e_2, \pm e_3, \frac{1}{2}(\pm 1 \pm e_1 \pm e_2 \pm e_3)\},$$

$$T' = \{\frac{1}{\sqrt{2}}(\pm 1 \pm e_1), \frac{1}{\sqrt{2}}(\pm e_2 \pm e_3), \frac{1}{\sqrt{2}}(\pm 1 \pm e_2), \frac{1}{\sqrt{2}}(\pm e_3 \pm e_1), \frac{1}{\sqrt{2}}(\pm 1 \pm e_3), \frac{1}{\sqrt{2}}(\pm e_1 \pm e_2)\}$$

(9)

where $T$ represents the elements of the binary tetrahedral group of order 24 and $T'$ with $T$ together represent the binary octahedral group. Either set represents the vertices of 24-cell, a platonic solid in 4D with the $W(F_4)$ symmetry of order 1152 [22]. Automorphism group of the root system of $A_3$ can be obtained via group extension technique by adjoining the Dynkin diagram symmetry generator $\gamma = [e_1, -e_1]^*$ to the group $W(A_3)$ which leads to the group

$$Aut(A_3) \approx W(B_3) \approx O_h = \{[p,\bar{p}] \oplus [p,\bar{p}]^* \oplus [t,\bar{t}] \oplus [t,\bar{t}]^*\}, \ p \in T, t \in T'. \tag{10}$$

Since we chose the roots in terms of imaginary quaternionic units then the group elements in (10) acting on the imaginary units can be further simplified as

$$Aut(A_3) \approx W(B_3) \approx O_h \approx S_4 \times C_2 = \{[p, \pm\bar{p}] \oplus [t, \pm\bar{t}]\}, \ p \in T, t \in T'. \tag{11}$$

The octahedral group $O_h$ in (10) can also be obtained from the Coxeter–Dynkin diagram of $B_3$ as shown in Figure 2. For further information about the finite subgroups of quaternions and the construction of the finite subgroups of $O(3)$ and $O(4)$ see the references [23, 24].

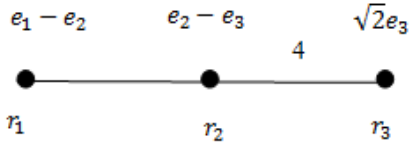

**Fig.2.** Coxeter diagram of $B_3$ with quaternionic simple roots

We can define the affine transformation in the root space of $A_3$ by introducing the affine reflection generators [25] in addition to the reflection generators in (7). However it suffices to introduce a translation generator in the direction of highest root. Let $\lambda$ be any root vector in the root system. Then the translation generator along the highest root $\alpha_H = -\alpha_0$ can be represented by

$$T(\lambda) = \lambda + \alpha_H . \tag{12}$$

Now the generators $r_1, r_2, r_3$ in (7) and $T$ generate the affine $W_a(A_3)$ in which the translations form an invariant Abelian subgroup. The root lattice $A_3$ is then the set of vectors $m_1\alpha_1 + m_2\alpha_2 + m_3\alpha_3$ where $m_i$ $(i=1,2,3)$ are integers. When expressed in terms of imaginary quaternionic units they represent the set of imaginary quaternionic units with integer coefficients. The root lattice is the sub-lattice in the weight lattice $A_3^*$ where a typical vector is defined by $a_1\omega_1 + a_2\omega_2 + a_3\omega_3 \equiv (a_1, a_2, a_3)$. Here $\omega_i$ $(i=1,2,3)$ defined in (4-5) constitute the basis vectors in the dual space of the root space and the coefficients $a_i$ $(i=1,2,3)$ are integers. The weight lattice consists of the set of imaginary quaternions with integer as well as half integer coefficients.

The sets of integers of the root and the weight lattices are related to each other by the Cartan matrix and its inverse as follows

$$\begin{bmatrix} a_1 \\ a_2 \\ a_3 \end{bmatrix} = \begin{bmatrix} 2 & -1 & 0 \\ -1 & 2 & -1 \\ 0 & -1 & 2 \end{bmatrix} \begin{bmatrix} m_1 \\ m_2 \\ m_3 \end{bmatrix}, \quad \begin{bmatrix} m_1 \\ m_2 \\ m_3 \end{bmatrix} = \frac{1}{4} \begin{bmatrix} 3 & 2 & 1 \\ 2 & 4 & 2 \\ 1 & 2 & 3 \end{bmatrix} \begin{bmatrix} a_1 \\ a_2 \\ a_3 \end{bmatrix}. \tag{13}$$

The first relation implies that for all integers $m_i, (i=1,2,3)$ the components $a_i, (i=1,2,3)$ take integer values but the second relation states that only certain subset of integer values of $a_i$, $m_i$ will take integer values. This explains also why the root lattice is a sub-lattice of the weight lattice.

The root system of $A_3$ can be written in terms of quaternionic unit vectors as

$$(\pm e_1 \pm e_2), (\pm e_2 \pm e_3), (\pm e_3 \pm e_1) . \tag{14}$$

It is customary in the crystallography books [26, 27] to use the vectors (now in the quaternionic notation) $e_1 + e_2$, $e_2 + e_3$, $e_3 + e_1$ as the generating basis vectors for the fcc lattice scaled by half of the lattice parameter. It is more appropriate to work in the weight lattice since it embeds the root lattice as a sub-lattice. The weight vectors $\omega_i$ $(i=1,2,3)$ are given in terms of quaternions as

$$\omega_1 = \frac{1}{2}(e_1 + e_2 + e_3), \quad \omega_2 = e_3, \quad \omega_3 = \frac{1}{2}(-e_1 + e_2 + e_3) . \tag{15}$$

The primitive vectors of the bcc lattice can also be taken as $\omega_1, \omega_1 - \omega_2, \omega_2 - \omega_3$ scaled by the lattice parameter. Orbits of $W(A_3)$ which correspond to the polyhedra [17] with tetrahedral/octahedral symmetry can be obtained from the "highest weight" vector [28] $\lambda = (a_1, a_2, a_3)$ where $a_i \geq 0, (i=1,2,3)$. We will use the notation $W(A_3)(a_1, a_2, a_3) \equiv (a_1, a_2, a_3)_{A_3}$ for the orbit representing the vertices of a certain polyhedron generated from the "highest weight" vector with $a_i \geq 0, (i = 1, 2, 3)$. For instance the orbits $(1,0,0)_{A_3}$, $(0,1,0)_{A_3}$, $(1,0,1)_{A_3}$, and $(1,1,1)_{A_3}$ represent the tetrahedron, octahedron, cuboctahedron (vertices represented by the non-zero roots of $A_3$), and the truncated octahedron respectively. One can find the vertices of the duals of these polyhedra represented by quaternions in a simple manner [29, 30]. The dual of the tetrahedron $(1,0,0)_{A_3}$, for example, is the tetrahedron represented by $(0,0,1)_{A_3}$. This implies that the tetrahedron is self-dual among the five Platonic solids, tetrahedron, octahedron, cube, icosahedron and dodecahedron.

The dual polyhedron of the cuboctahedron $(1,0,1)_{A_3}$ is the rhombic dodecahedron (Wigner-Seitz cell or Voronoi cell in general) whose vertices are represented by the union of the orbits $(1,0,0)_{A_3} \oplus (0,1,0)_{A_3} \oplus (0,0,1)_{A_3}$ [30]. The weight lattice $A_3^*$ is the bcc lattice whose Voronoi cell is the orbit $(1,1,1)_{A_3}$ which represents the 24 vertices of the truncated octahedron. The highest weight $(1,1,1)$ here is known as the Weyl vector. Dual of the truncated octahedron is the Catalan solid (tetrakis hexahedron) whose vertices are represented by the union of the orbits $(1,0,0)_{A_3} \oplus \frac{3}{4}(0,1,0)_{A_3} \oplus (0,0,1)_{A_3}$ [30]. However, the tetrakis hexahedron does not describe the bcc lattice in the direct space rather it is the cube which is represented by the union two tetrahedra $(1,0,0)_{A_3} \oplus (0,0,1)_{A_3}$ in our notation which describes the first nearest points of bcc lattice and the next nearest lattice points are represented by the six vertices of the octahedron $(0,1,0)_{A_3}$.

There are three maximal subgroups of the octahedral group $Aut(A_3) \approx W(B_3)$, namely, the tetrahedral group $W(A_3)$, the chiral octahedral group consisting of the elements $W(B_3)/C_2 = \{[p, \bar{p}] \oplus [t, \bar{t}]\}$ and the pyritohedral group consisting of the elements $T_h \approx A_4 \times C_2 = \{[p, \pm \bar{p}]\}$. The pyritohedral symmetry represents the symmetry of the pyritohedron which is an irregular dodecahedron with irregular pentagonal faces that occurs in pyrites.

## 3. Construction of the affine Coxeter group $W_a(A_4)$ in terms of quaternions

It is now straightforward to construct the root and weight lattices of the $A_n, (n = 1, 2, 3, 4, 5, ...)$ series of the Coxeter-Weyl groups. The affine group $W_a(A_1)$ is the symmetry of the one dimensional lattice. The affine group $W_a(A_2)$ describes the symmetry of the honeycomb lattice; the best example is the graphene. Now we want to discuss the $A_4$ lattices whose projections onto the Coxeter plane describe the decagonal quasicrystal structures.

The Cartan matrix and its inverse matrix are given by the respective matrices as follows

$$C_{A_4} = \begin{pmatrix} 2 & -1 & 0 & 0 \\ -1 & 2 & -1 & 0 \\ 0 & -1 & 2 & -1 \\ 0 & 0 & -1 & 2 \end{pmatrix}, (C_{A_4})^{-1} = \frac{1}{5}\begin{pmatrix} 4 & 3 & 2 & 1 \\ 3 & 6 & 4 & 2 \\ 2 & 4 & 6 & 3 \\ 1 & 2 & 3 & 4 \end{pmatrix}. \tag{16}$$

The automorphism group $Aut(A_4) \approx W(A_4):\gamma$ of the root system of $A_4$ where $\gamma$ is the Dynkin diagram symmetry has an elegant representation in terms of icosians $I$ (quaternionic elements of the binary icosahedral group) and its algebraic conjugate $\tilde{I}$. This follows by choosing the simple roots of $A_4$ from the set of icosians $I$. Sometime back we have constructed the Coxeter–Weyl group $W(A_4)$ in terms of quaternions as one of the maximal subgroup of the Coxeter group $W(H_4)$ [18] and studied the properties of its quasi-regular polytopes and their dual polytopes [20]. Let the simple roots of $A_4$ be given by the following quaternions

$$\alpha_1 = -\sqrt{2}, \alpha_2 = \frac{1}{\sqrt{2}}(1+e_1+e_2+e_3), \alpha_3 = -\sqrt{2}e_1, \alpha_4 = \frac{1}{\sqrt{2}}(e_1 - \sigma e_2 - \tau e_3). \tag{17}$$

Here $\tau = \frac{1+\sqrt{5}}{2}, \sigma = \frac{1-\sqrt{5}}{2}$ satisfy the relations $\tau\sigma = -1, \tau + \sigma = 1, \tau^2 = \tau + 1$, and $\sigma^2 = \sigma + 1$. The elements of the group $W(A_4)$ can be compactly written in the form

$$W(A_4) = \{[p, -c\bar{\tilde{p}}c] \oplus [p, c\bar{\tilde{p}}c]^*\}. \tag{18}$$

Here $p \in I$ is an arbitrary element of the binary icosahedral group $I$ with $c = \frac{1}{\sqrt{2}}(e_2 - e_3)$ and $\tilde{p} = p(\tau \leftrightarrow \sigma)$ is an element of the representation of the binary icosahedral group $\tilde{I}$ obtained from $I$ by interchanging $\tau$ and $\sigma$. The extended Coxeter-Dynkin diagram playing a crucial role in this section is depicted in Figure 3.

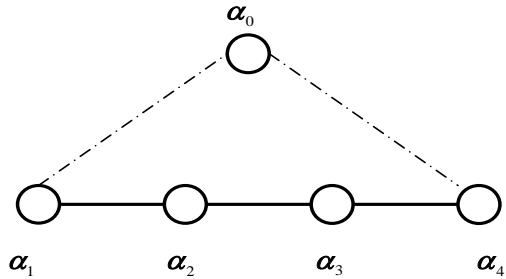

**Fig.3**. Extended Coxeter-Dynkin diagram of $A_4$

The automorphism point group of the $A_4$ lattice is given as

$$Aut(A_4) \approx W(A_4):\gamma = \{[p, \mp c\bar{\tilde{p}}c] \oplus [p, \pm c\bar{\tilde{p}}c]^*\}. \tag{19}$$

The element $c = \frac{1}{\sqrt{2}}(e_2 - e_3)$ belongs to the set of quaternions $T'$ given in (9). The basis vectors of the weight lattice and the highest root are given in terms of quaternions as follows

$$\omega_1 = \frac{1}{\sqrt{10}}(-\sqrt{5} + \tau e_2 - \sigma e_3), \qquad \omega_2 = \frac{1}{\sqrt{10}}(2\tau e_2 - 2\sigma e_3),$$

$$\omega_3 = \frac{1}{\sqrt{10}}(-\sqrt{5}e_1 + \tau^2 e_2 - \sigma^2 e_3), \qquad \omega_4 = \frac{1}{\sqrt{10}}(2e_2 - 2e_3) \equiv -\frac{2}{\sqrt{5}}c,$$

$$\alpha_H = \alpha_1 + \alpha_2 + \alpha_3 + \alpha_4 = \omega_1 + \omega_4 = \frac{1}{\sqrt{2}}(-1 + \tau e_2 + \sigma e_3). \tag{20}$$

### 3.1 The root lattice $A_4$

The root lattice can be generated by using the generators of the affine Coxeter-Weyl group $W_a(A_4)$ given as follows

$$r_1 = [1, -1]^*, r_2 = [\frac{1}{2}(1 + e_1 + e_2 + e_3), -\frac{1}{2}(1 + e_1 + e_2 + e_3)]^*, r_3 = [e_1, -e_1]^*,$$

$$r_4 = [\frac{1}{2}(e_1 - \sigma e_2 - \tau e_3), -\frac{1}{2}(e_1 - \sigma e_2 - \tau e_3)]^*, T(\lambda) = \lambda + \alpha_H. \tag{21}$$

The root lattice consists of the set of vectors $m_1\alpha_1 + m_2\alpha_2 + m_3\alpha_3 + m_4\alpha_4$ with $m_i (i = 1, 2, 3, 4)$ integers. However it is customary to work in the weight lattice which admits the root lattice as a sub-lattice where any weight lattice vector is given by $a_1\omega_1 + a_2\omega_2 + a_3\omega_3 + a_4\omega_4$ with $a_i (i = 1, 2, 3, 4)$ integers. Relations similar to those given in (13) exist where the Cartan matrix and its inverse in (16) must be used. The polytope $(1, 0, 0, 0)_{A_4}$ (5-cell) has 5 vertices, 10 triangular faces, 10 edges and 5 tetrahedral facets (4-cells) [20]. The polytope $(0, 1, 0, 0)_{A_4}$ (rectified 5-cell) is an Archimedean solid in 4D and possesses 10 vertices, 30 edges, 30 triangular faces and 10 facets (5 tetrahedra+5 octahedra). The structure of the polytope $(0, 0, 1, 0)_{A_4}$ consists of vectors with the negatives of the vectors of the polytope $(0, 1, 0, 0)_{A_4}$ since it is obtained from the rectified 5-cell by the Dynkin diagram symmetry operator. The same is also true for the polytope $(0, 0, 0, 1)_{A_4}$ which is a 5-cell obtained by the Dynkin diagram symmetry, moreover it is the dual polytope of the 5-cell $(1, 0, 0, 0)_{A_4}$; it is this property that 5-cell is said to be self-dual. As we will see later, when one of these sets of vectors of 5-cell is projected onto the Coxeter plane, they play the role of the de Brujin's parameter $\zeta = \exp(2\pi i / 5)$ [8]. The non-zero 20 roots, compactly denoted in the weight lattice notation by the orbit $W(A_4)(1, 0, 0, 1) \equiv (1, 0, 0, 1)_{A_4}$, is a polytope possessing 20 vertices, 60 edges, 70 faces (40 equilateral triangles+30 squares) and 30 facets (10 tetrahedra + 20 triangular prisms) which is the extension of the cuboctahedron to 4D. Therefore the polytope $(1, 0, 0, 1)_{A_4}$ in 4D plays the same role as the cuboctahedron plays in 3D. If we follow the arguments raised in Section 2 the Voronoi cell of the root lattice $A_4$ is the dual polytope of the polytope $(1,0,0,1)_{A_4}$. It can be shown that the dual polytope, that is to say, the Voronoi cell of the root lattice is the union of the orbits [20]

$$(1,0,0,0)_{A_4} \oplus (0,1,0,0)_{A_4} \oplus (0,0,1,0)_{A_4} \oplus (0,0,0,1)_{A_4}. \tag{22}$$

The Voronoi cell of the root lattice consists of 30 vertices, 70 edges, 60 faces (all rhombuses) and 20 facets (all rhombohedra). The typical cell, the rhombohedron of the Voronoi cell of the root lattice is shown in Figure 4. Its projection onto the Coxeter plane is very important for the study of the decagonal quasicrystals as we will discuss in the next section.

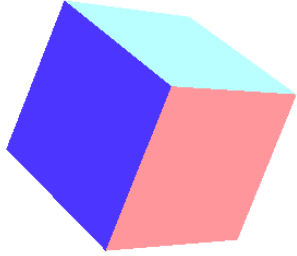

**Fig.4.** A typical cell of the Voronoi polytope of the root lattice of $A_4$

### 3.2 The weight lattice $A_4^*$

Following the discussions of the weight lattice of $A_3^*$ we easily recognize the Voronoi cell of the weight lattice $A_4^*$ as the polytope $(1,1,1,1)_{A_4}$. It consists of 120 vertices, 240 edges, 150 faces (60 equilateral triangles+90 squares) and 30 facets (10 truncated octahedral +20 hexagonal prisms). The dual polytope of the Voronoi cell $(1,1,1,1)_{A_4}$ is the polytope which is the union of the orbits [20]

$$(1,0,0,0)_{A_4} \oplus \frac{2}{3}(0,1,0,0)_{A_4} \oplus \frac{2}{3}(0,0,1,0)_{A_4} \oplus (0,0,0,1)_{A_4}. \tag{23}$$

It consists of 30 vertices, 150 edges, 240 faces (it consists of two types of scalene triangles) and 20 solids as facets consisting of four vertices and four triangles of two kind scalene triangles as shown in Figure5.

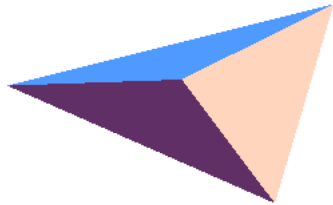

**Fig.5.** A typical facet of the dual polytope of the Voronoi cell of the weight lattice $A_4^*$

# 4. Orthogonal projection of the $A_4$ lattices onto the Coxeter plane and the decagonal quasicrystallography

We first give a general projection technique before we apply it to the $A_4$ lattices.

## 4.1 Orthogonal projection onto the Coxeter plane

In this section we will develop a general technique for the orthogonal projection of an arbitrary lattice vector onto the Coxeter plane. For further details see for instance the reference [31]. We will illustrate the cases for the simply-laced root systems. Let the rank-$n$ Coxeter-Weyl group be given by $W(G) = \langle r_1, r_2, ..., r_n \rangle$ with the simple roots $\alpha_i, (i = 1, ..., n)$. Let us assume that the odd and even numbered roots are partitioned into orthogonal sets. This is certainly true for the $A_n$ series with even $n$ and similar decompositions can be arranged for the rest of the Coxeter-Weyl groups. Let us define the generators [32] $R_1 = r_1 r_3 ... r_{n-1}, R_2 = r_2 r_4 ... r_n$ and the roots

$$\beta_1 = x_1 \alpha_1 + x_3 \alpha_3 + ... + x_{n-1} \alpha_{n-1}, \beta_2 = x_2 \alpha_2 + x_4 \alpha_4 + ... + x_n \alpha_n. \tag{24}$$

Without loss of generality, the Coxeter element of the group $W(G)$ can be taken as $R_1 R_2$ satisfying the relation $(R_1 R_2)^h = 1$. Here $h$ represents the Coxeter number. One can prove that the coefficients $x_i (i = 1, ..., n)$ can be determined as the components of the eigenvector of the incidence matrix $(2I - C_G)X = cX$ where $c = 2\cos\frac{\pi}{h}$ is the largest eigenvalue of the incidence matrix. Because of the Perron-Frobenius theorem the components of the eigenvector $X$ are all real and positive. Since the eigenvectors are determined up to normalization, the coefficients $x_i (i = 1, ..., n)$ are determined from the norm fixing of the roots $(\beta_1, \beta_1) = (\beta_2, \beta_2) = 2$. This implies the normalization $\sum_{i=1,3,} x_i x_i = \sum_{i=2,4,} x_i x_i = 1$. Then one can prove that $R_1$ and $R_2$ act on the roots as reflection generators where the roots satisfy $(\beta_1, \beta_2) = -c$. The group generated by $R_1$ and $R_2$ is the maximal dihedral subgroup $D_h$ of the Coxeter group $W(G)$ and it is the noncrystallographic group represented by the Coxeter diagram $I_2(h)$ for $h \neq 2, 3, 4, 6$. The plane determined by the roots $\beta_1$ and $\beta_2$ is defined as the Coxeter plane. An orthogonal set of unit vectors can be defined in the Coxeter plane, for example, by

$$\hat{x} = \frac{1}{\sqrt{2(2+c)}} (\beta_2 - \beta_1), \hat{y} = \frac{1}{\sqrt{2(2-c)}} (\beta_2 + \beta_1) \ . \tag{25}$$

Then a vector of the weight lattice $\lambda = \sum_{i=1}^{n} a_i \omega_i$ can be projected by the formula

$$\lambda_x = \frac{1}{\sqrt{2(2-c)}}[-(x_1a_1 + x_3a_3 + ...) + (x_2a_2 + x_4a_4 + ...)]$$

$$\lambda_y = \frac{1}{\sqrt{2(2+c)}}[(x_1a_1 + x_3a_3 + ...) + (x_2a_2 + x_4a_4 + ...)].$$
(26)

## 4.2 Orthogonal projection of the $A_4$ lattices onto the Coxeter plane

Since the Coxeter number of the group $A_4$ is given by $h = 5$ the corresponding dihedral symmetry $D_5$ is generated by the Coxeter graph $H_2$ where the defining graph is shown in Figure 6 with the simple roots $\beta_i$ $(i = 1, 2)$ and the reflection generators $R_i$ $(i = 1, 2)$.

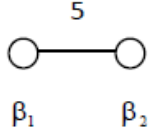

**Fig.6.** Coxeter diagram of $H_2$

One can write the respective generators and the simple roots of the Coxeter group $W(H_2) \approx D_5$ in terms of the reflection generators and the simple roots of $W(A_4)$ as follows

$$R_1 = r_1 r_3, R_2 = r_2 r_4, \ \beta_a = x_{ai}\alpha_i, \ (a = 1, 2), (i = 1, 2, 3, 4).$$
(27)

Using (26) we determine the simple roots of $H_2$ as

$$\beta_1 = \frac{1}{\sqrt{2+\tau}}(\alpha_1 + \tau\alpha_3), \beta_2 = \frac{1}{\sqrt{2+\tau}}(\tau\alpha_2 + \alpha_4).$$
(28)

Note that the Dynkin diagram symmetry of the Coxeter-Dynkin diagram of $A_4$ implies the Coxeter diagram symmetry of the $H_2$ diagram. Had we chosen the angles between the first and the second roots of the Coxeter graph of $H_2$ as $72^0$ corresponding to the symbol 5/3 instead of 5 in Figure 6 then a similar root system could have been obtained as

$$\beta_1' = \frac{1}{\sqrt{2+\sigma}}(\alpha_1 + \sigma\alpha_3), \beta_2' = \frac{1}{\sqrt{2+\sigma}}(\sigma\alpha_2 + \alpha_4).$$
(29)

Actually the root vectors $\beta_1, \beta_2, \beta_1', \beta_2'$ decompose the four dimensional Euclidean space into two 2D orthogonal spaces. Here the scalar products are given as

$$(\beta_1, \beta_1) = (\beta_2, \beta_2) = (\beta_1', \beta_1') = (\beta_2', \beta_2') = 2$$

$$(\beta_1, \beta_2) = -\tau, (\beta_1', \beta_2') = -\sigma, (\beta_i, \beta_j') = 0, (i, j = 1, 2).$$
(30)

This follows also from the fact that the Cartan matrix and its inverse in (16) can be block-diagonalized in terms of $2\times 2$ matrices corresponding to two different representations of the graph $H_2$. To complete this let us define $\beta_3 \equiv \beta_1', \beta_4 \equiv \beta_2'$ and convert the coefficients in (29-30) into a matrix then the block diagonalized Cartan matrix turns out to be

$$XC_{A_4}X^T = C'_{A_4}, \quad C'_{A_4} = \begin{pmatrix} 2 & -\tau & 0 & 0 \\ -\tau & 2 & 0 & 0 \\ 0 & 0 & 2 & -\sigma \\ 0 & 0 & -\sigma & 2 \end{pmatrix}. \tag{31}$$

The above technique can be applied to any root or weight lattices. We will discuss the projection of the $A_4$ lattices onto the plane $E_\parallel$ defined by the vectors $\beta_1$ and $\beta_2$. The plane determined by the vectors $\beta_3$ and $\beta_4$ is usually denoted by $E_\perp$. For orthogonal projections we have to define orthogonal unit vectors in the planes $E_\parallel$ and $E_\perp$ as follows

$$\hat{x} = \frac{1}{\sqrt{2(2+\tau)}}(-\beta_1 + \beta_2), \quad \hat{y} = \frac{\tau}{\sqrt{2}}(\beta_1 + \beta_2),$$

$$\hat{z} = \frac{1}{\sqrt{2(2+\sigma)}}(-\beta_3 + \beta_4), \quad \hat{w} = \frac{\sigma}{\sqrt{2}}(\beta_1 + \beta_2). \tag{32}$$

In terms of the simple roots of $A_4$ they read

$$\hat{x} = \frac{1}{(2+\tau)\sqrt{2}}(-\alpha_1 + \tau\alpha_2 - \tau\alpha_3 + \alpha_4), \quad \hat{y} = \frac{\tau}{\sqrt{2(2+\tau)}}(\alpha_1 + \tau\alpha_2 + \tau\alpha_3 + \alpha_4)$$

$$\hat{z} = \frac{1}{(2+\sigma)\sqrt{2}}(-\alpha_1 + \sigma\alpha_2 - \sigma\alpha_3 + \alpha_4), \quad \hat{w} = \frac{\sigma}{\sqrt{2(2+\sigma)}}(\alpha_1 + \sigma\alpha_2 + \sigma\alpha_3 + \alpha_4). \tag{33}$$

Now the components of an arbitrary weight lattice vector $\lambda = a_1\omega_1 + a_2\omega_2 + a_3\omega_3 + a_4\omega_4$ will be given by

$$\lambda_x = \frac{1}{(2+\tau)\sqrt{2}}[-a_1 + a_4 + \tau(a_2 - a_3)], \quad \lambda_y = \frac{\tau}{\sqrt{2(2+\tau)}}[a_1 + a_4 + \tau(a_3 + a_2)],$$

$$\lambda_z = \frac{1}{(2+\sigma)\sqrt{2}}[-a_1 + a_4 + \sigma(a_2 - a_3)], \quad \lambda_w = \frac{\sigma}{\sqrt{2(2+\sigma)}}[a_1 + a_4 + \sigma(a_3 + a_2)]. \tag{34}$$

The components of an arbitrary vector $\Lambda = m_1\alpha_1 + m_2\alpha_2 + m_3\alpha_3 + m_4\alpha_4$, with $m_i \in \mathbf{Z}$ in the root lattice can be determined as

$$\Lambda_x = \frac{1}{\sqrt{2}}(-m_1 + \tau m_2 - \tau m_3 + m_4), \quad \Lambda_y = \frac{1}{\sqrt{2(\tau+2)}}(-\sigma m_1 + m_2 + m_3 - \sigma m_4),$$

$$\Lambda_z = \frac{1}{\sqrt{2}}(-m_1 + \sigma m_2 - \sigma m_3 + m_4), \quad \Lambda_w = \frac{1}{\sqrt{2(\sigma+2)}}(-\tau m_1 + m_2 + m_3 - \tau m_4). \tag{35}$$

The generators of the group $W(H_2) \approx D_5$ acting on the root vectors $\beta_1, \beta_2, \beta_3,$ and $\beta_4$ would read

$$R_1 = \begin{pmatrix} -1 & 0 & 0 & 0 \\ \tau & 1 & 0 & 0 \\ 0 & 0 & -1 & 0 \\ 0 & 0 & \sigma & 1 \end{pmatrix}, \quad R_2 = \begin{pmatrix} 1 & \tau & 0 & 0 \\ 0 & -1 & 0 & 0 \\ 0 & 0 & 1 & \sigma \\ 0 & 0 & 0 & -1 \end{pmatrix}. \tag{36}$$

The same generators take the forms in the orthogonal bases $\hat{x}, \hat{y}, \hat{z},$ and $\hat{w}$

$$R_1' = \frac{1}{2}\begin{pmatrix} -\tau & \sqrt{2+\sigma} & 0 & 0 \\ \sqrt{2+\sigma} & \tau & 0 & 0 \\ 0 & 0 & -\sigma & \sqrt{2+\tau} \\ 0 & 0 & \sqrt{2+\tau} & \sigma \end{pmatrix}, \quad R_2' = \frac{1}{2}\begin{pmatrix} -\tau & -\sqrt{2+\sigma} & 0 & 0 \\ -\sqrt{2+\sigma} & \tau & 0 & 0 \\ 0 & 0 & -\sigma & -\sqrt{2+\tau} \\ 0 & 0 & -\sqrt{2+\tau} & \sigma \end{pmatrix}.$$

$$\tag{37}$$

In the orthogonal basis of the unit vectors $\hat{x}, \hat{y}, \hat{z},$ and $\hat{w}$ the rotation matrix will take a simple form

$$R_1'R_2' = \begin{pmatrix} \cos\frac{2\pi}{5} & \sin\frac{2\pi}{5} & 0 & 0 \\ -\sin\frac{2\pi}{5} & \cos\frac{2\pi}{5} & 0 & 0 \\ 0 & 0 & \cos\frac{4\pi}{5} & \sin\frac{4\pi}{5} \\ 0 & 0 & -\sin\frac{4\pi}{5} & \cos\frac{4\pi}{5} \end{pmatrix}. \tag{38}$$

This corresponds to the rotation $72^0$ in the plane $E_\parallel$ and the rotation $144^0$ in the plane $E_\perp$. The point group $W(A_4)$ acting on a general vector $\lambda = a_1\omega_1 + a_2\omega_2 + a_3\omega_3 + a_4\omega_4$ would generate a set of 120 vectors. A list of vectors $W(A_4)(a_1\omega_1 + a_2\omega_2 + a_3\omega_3 + a_4\omega_4)$ is given in Appendix A expressed in terms of the orbits of the dihedral group $D_5 \approx W(H_2)$. As an example let us study the projections of the 5-cells $(1,0,0,0)_{A_4}$ and $(0,0,0,1)_{A_4}$ onto the plane $E_\parallel$. From Appendix A we obtain the vertices of the 5 cell in 4D that are given by the vectors as $(1,0,0,0), (-1,1,0,0), (0,0,-1,1), (0,0,0,-1), (0,-1,1,0)$. When they are projected onto the Coxeter plane $E_\parallel$ they can be represented in terms of de Brujin's parameter $\zeta = (\cos(\frac{2\pi}{5}), \sin(\frac{2\pi}{5}))$ as

$$(1,0,0,0) \to \sqrt{\tfrac{2}{5}}\zeta = \tfrac{1}{2}\sqrt{\tfrac{2}{5}}(-\sigma, \sqrt{2+\tau}), \qquad (-1,1,0,0) \to \sqrt{\tfrac{2}{5}}\zeta^2 = \tfrac{1}{2}\sqrt{\tfrac{2}{5}}(-\tfrac{\tau}{2}, \tfrac{\sqrt{2+\sigma}}{2}),$$

$$(0,0,-1,1) \to \sqrt{\tfrac{2}{5}}\zeta^3 = \tfrac{1}{2}\sqrt{\tfrac{2}{5}}(-\tfrac{\tau}{2}, -\tfrac{\sqrt{2+\sigma}}{2}), \quad (0,0,0,-1) \to \sqrt{\tfrac{2}{5}}\zeta^4 = \tfrac{1}{2}\sqrt{\tfrac{2}{5}}(-\tfrac{\sigma}{2}, -\tfrac{\sqrt{2+\tau}}{2}),$$

$$(0,-1,1,0) \to \sqrt{\tfrac{2}{5}}\zeta^0 = \sqrt{\tfrac{2}{5}}(1,0).$$

(39)

The vertices of the projected 5-cell $(0,0,0,1)_{A_4}$ are the negatives of those in (39). The projected 5-cells are illustrated in Figure 7(a-b). Note that Robinson's triangles naturally occur in the projected 5-cells. It is also interesting to observe that the two polytopes $(1,0,0,0)_{A_4}$ and $(0,0,1,0)_{A_4}$ represent the weights of the $\underline{5}$ and $\underline{10}^*$ representations of the $SU(5)$ Lie group used as a model for the Grand Unified Theory (GUT) in particle physics [33]. The projections of the polytopes $(0,1,0,0)_{A_4}$ and $(0,0,1,0)_{A_4}$ are given in Figure 8(a-b).

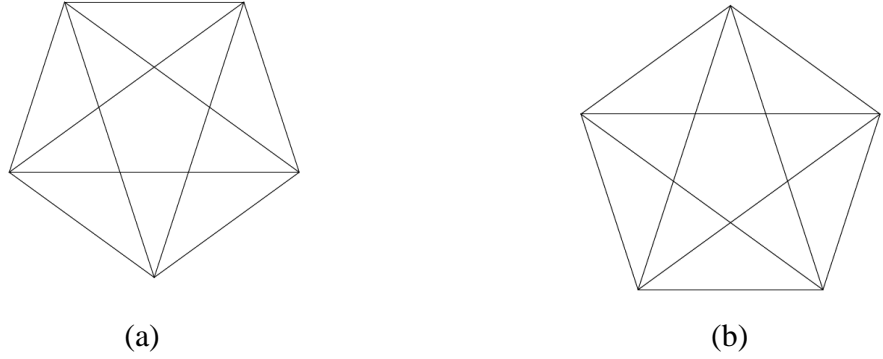

(a)        (b)

**Fig.7**. Projected 5-cells (a) the polytope $(1,0,0,0)_{A_4}$ and (b) the polytope $(0,0,0,1)_{A_4}$

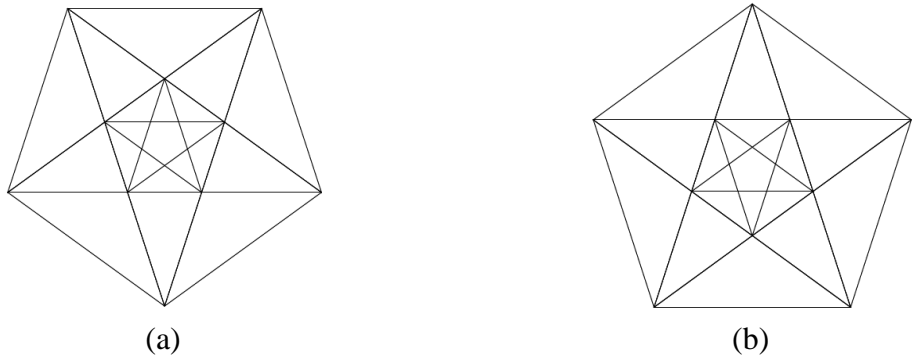

(a)        (b)

**Fig.8**. Projected polytopes (a) $(0,0,1,0)_{A_4}$ and (b) $(0,1,0,0)_{A_4}$

We will not discuss the projections of all quasi regular polytopes of $W(A_4)$ except those which seem to be related to the structures obtained by the projection of the cube $Q_5$ in 5D onto the Coxeter plane. The five dimensional cube has 32 vertices and can be obtained as a polytope from the Coxeter-Dynkin diagram $B_5$ (see Figure 9) by acting $W(B_5)$ on the highest weight vector $(0,0,0,0,1)$.

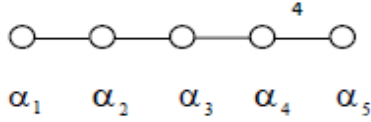

**Fig.9**. Coxeter-Dynkin diagram of $B_5$

Therefore the 5D cube, that is, one of the regular polytope of $B_5$ in our notation will read $W(B_5)(0,0,0,0,1) \equiv (0,0,0,0,1)_{B_5}$. The group $W(A_4)$ is a maximal subgroup in the group $W(B_5)$ and has an index equal to 32 which represents the number of vertices of the cube $Q_5$. This follows from the fact that $W(A_4)$ leaves the $B_5$ vector $(0,0,0,0,1)$ invariant. Note also that the group $W(B_5)$ has the Coxeter number $h=10$ so that it has a maximal dihedral symmetry $D_{10}$ acting in the Coxeter plane. The branching, in other words, the decomposition of the cube $Q_5$ under the group $W(A_4)$ can be written as

$$(0,0,0,0,1)_{B_5} = 2(0,0,0,0)_{A_4} \oplus (1,0,0,0)_{A_4} \oplus (0,1,0,0)_{A_4} \oplus (0,0,1,0)_{A_4} \oplus (0,0,0,1)_{A_4}. \quad (40)$$

In particle physicists' notation it is the decomposition of the 32 dimensional spinor representation of the $SO(11)$ Lie group under the $SU(5)$ as [34]

$$\underline{32} = 2(\underline{1}) + \underline{5} + \underline{10}^* + \underline{10} + \underline{5}^*. \quad (41)$$

After this little digression which will be very useful in the following discussions we continue studying the projections of the polytopes corresponding to the cells of the root and weight lattices. The projection of the polytope $(1,0,0,1)_{A_4}$ can be carried by determining its 20 vertices using Appendix A and projecting them onto the Coxeter plane by using (34). The projected point distributions and the projected polytope are given in Figure 10.

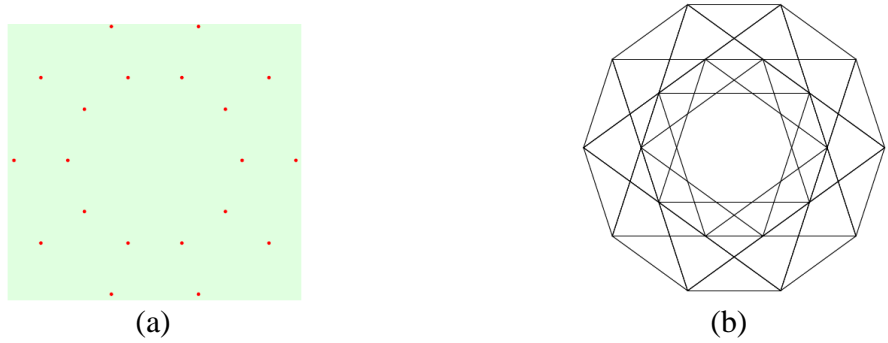

**Fig.10.** The root system of $A_4$ projected onto the Coxeter plane
(a) root system (b) the polytope

The projection of the vertices of the five dimensional cube $Q_5$ can be found in the literature, for instance (see M. Senechal in ref. [4], p.61, Fig. 2.12). There are two points at the center and the other 30 points are on three circles of radii varying proportional to the golden ratio $\tau = \dfrac{1+\sqrt{5}}{2}$. Let us compare the set of vertices of the Voronoi cell of the root lattice given in (22) and the decomposition of the $Q_5$ in (40) and (41). It is clear that when the doubly degenerate central point of the projected figure of $Q_5$ is removed then the rest will display the projection of the Voronoi cell of the root lattice of $A_4$ which is shown in Figure 11.

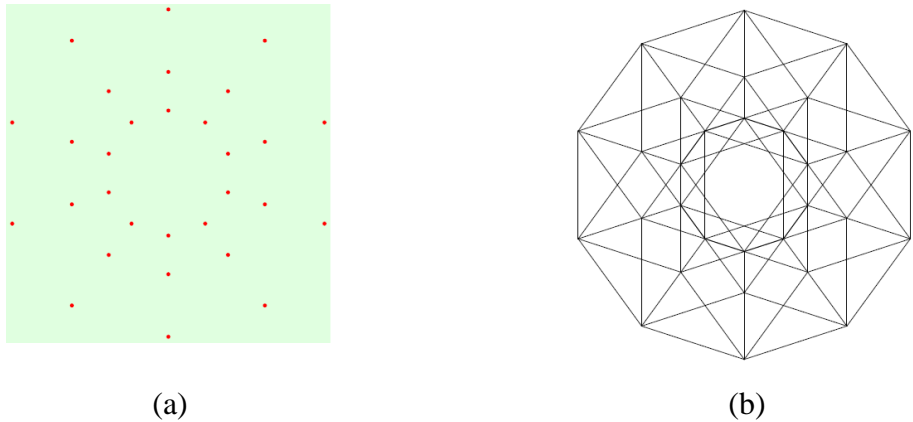

**Fig.11.** The orthogonal projection of the Voronoi cell of the root lattice onto the Coxeter plane (a) points (b) dual of the polytope (1,0,0,1)

Let us remember that the facets of the Voronoi cells here are the rhombohedra as was shown in Figure 4. The work of R. A. Al Ajlouni [14] has pointed out that the ancient Muslim designers were able to construct a variety of quasi-periodic patterns. After examining a large number of Islamic patterns she concludes that the underlying basic structure, invariant under 10-fold rotations, is the quasi-periodic cartwheel pattern just like the one which we illustrated in Figure 11. It is exactly the nested decagrams growing as the power of the golden ratio $\tau^n$. Indeed, a quasi lattice obtained by projection involves many other points in addition to the a whole pattern growing with $\tau^n$.

Now we discuss the orthogonal projection of the Voronoi cell of the $A_4^*$ lattice. The lattice points, in other words, the 120 vertices of the Voronoi cell of the weight lattice $(1,1,1,1)_{A_4}$ can be obtained from Appendix A. When they are orthogonally projected onto the Coxeter plane using (33) we obtain the point distributions in Figure 12. This mapping of the Voronoi cell of the lattice $A_4^*$ has been made by Baake et, al. [12] by a different method.

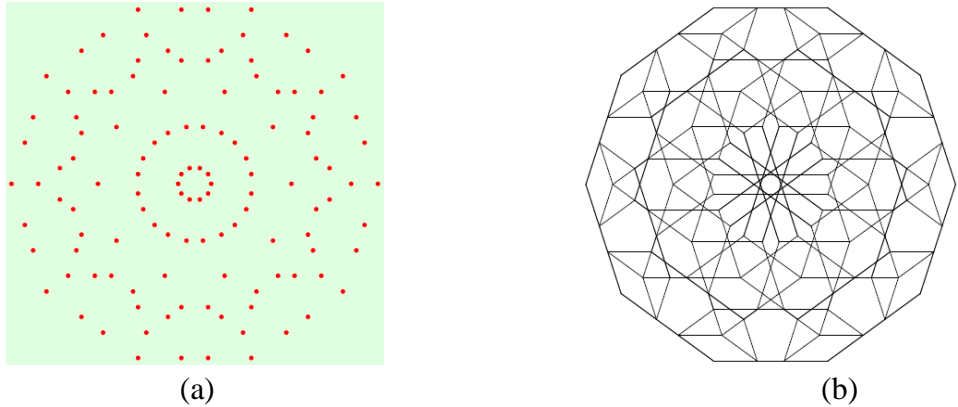

(a)          (b)

**Fig. 12**. Orthogonal projection of the polytope $(1,1,1,1)_{A_4}$
(a) point distribution (b) the Voronoi cell of the weight lattice

Before concluding we point out a few more interesting results. The vertices of the polytope $(0,1,1,0)_{A_4}$ can also be obtained from the root vectors since the highest weight satisfies $\omega_2 + \omega_3 = \alpha_1 + 2\alpha_2 + 2\alpha_3 + \alpha_4$. This polytope has 30 vertices, 60 edges, 40 faces (20 equilateral triangles+20 regular hexagons) and 10 facets (all truncated tetrahedra). Its orthogonal projection is depicted in Figure 13. It has been argued by Senechal [17] (p.201, Fig.6.23) that it corresponds to a generalized Penrose tiling. The dual of this polytope is represented as the union of two 5-cells $(1,0,0,0)_{A_4} \oplus (0,0,0,1)_{A_4}$. The facet of the dual polytope is a solid as shown in Figure 14 with 4 vertices, 4 faces (consisting of isosceles triangles) with a Klein four-group symmetry $C_2 \times C_2$ of order 4.

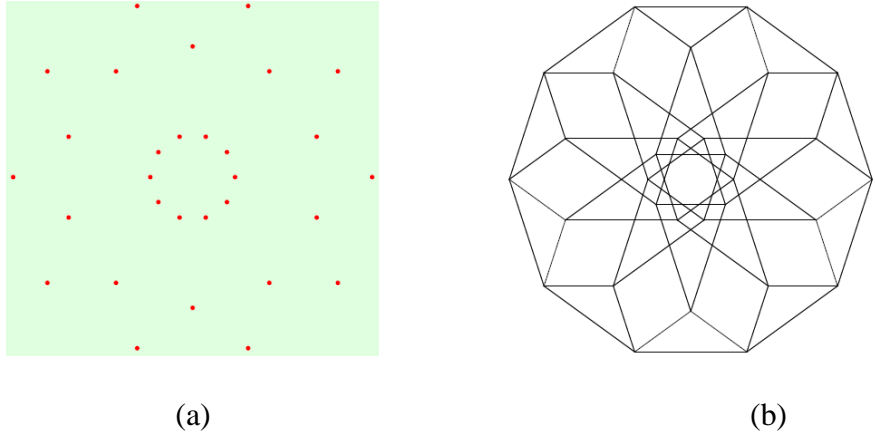

(a)                      (b)

**Fig.13**. Orthogonal projection of the polytope $(0,1,1,0)_{A_4}$
(a) point distribution (b) the polytope

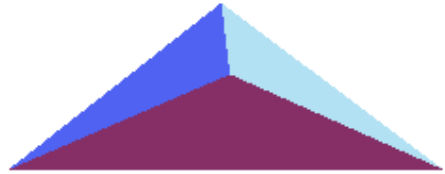

**Fig.14**. A typical facet of the polytope $(1,0,0,0)_{A_4} \oplus (0,0,0,1)_{A_4}$

The strip projections of the $A_4$ lattices will be discussed below. The Voronoi cells of the root and weight lattices define circles on the plane $E_\perp$ as shown in Figure 11 and Figure 12. One cuts the $A_4$ lattice by a cylinder based on the circle defined by the projection of the Voronoi cell onto the plane $E_\perp$. The lattice points which remain in the cylinder are to be projected onto the Coxeter plane $E_\parallel$. This is the strip projection which derives its name from the projection of the square lattice onto a line with a slope $-\sigma$ and with a limited lattice points constrained by two parallel lines. The distributions of the points for the root lattice obtained from the strip projection are shown in Figure 15.

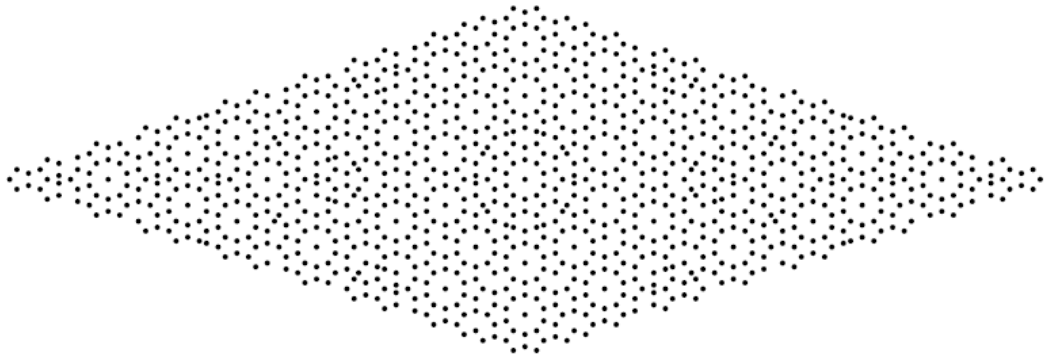

**Fig.15**. The quasicrystal obtained by projecting the root lattice onto the Coxeter plane

By replacing the cylinder in the root lattice with the cylinder based on the circle represented by the projection of the Voronoi cell $(1,1,1,1)_{A_4}$ of the weight lattice one determines the set of lattice points to be projected onto the Coxeter plane $E_\parallel$. The quasicrystal obtained by this projection is displayed in Figure 16.

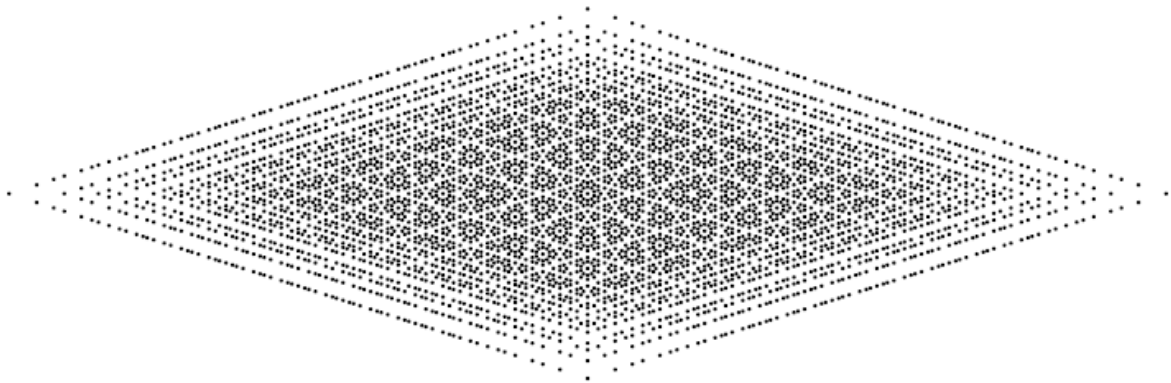

**Fig.16**. The quasicrystal obtained by strip projection of the weight lattice onto the Coxeter plane

For decorative purposes below we display some point distributions which are not obtained from the strip projections. They are given in Figure 17.

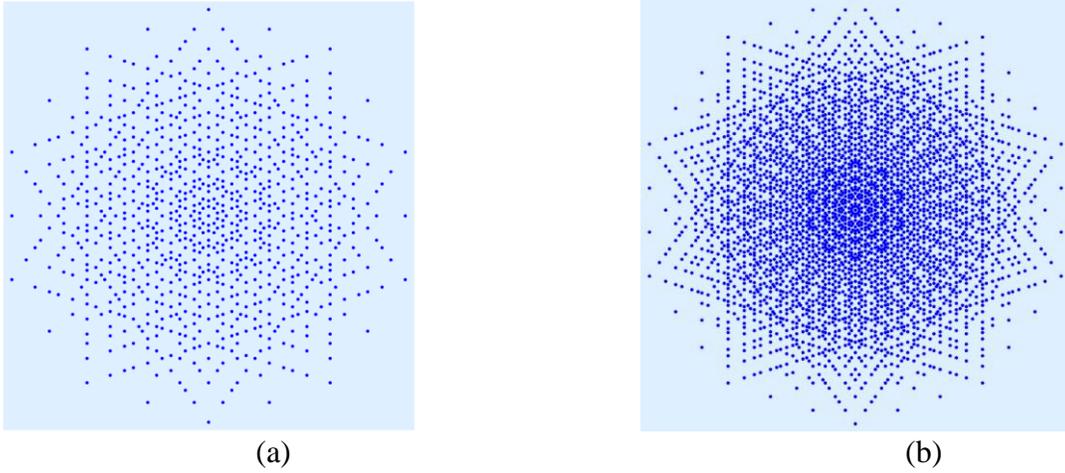

             (a)                                  (b)

**Fig. 17**. Decagram distributions involving some decorative points.
$(a)$ constructed from the set of points $-2 \leq |a_i| \leq 2, \ i = 1, 2, 3, 4; \ \ |a_1 + a_2 + a_3 + a_4| \leq 2,$
$(b)$ constructed from the set of points $-3 \leq |a_i| \leq 3, \ i = 1, 2, 3, 4; \ \ |a_1 + a_2 + a_3 + a_4| \leq 3$

## 5. Conclusion

We have pointed out a number of interesting structures of the $A_4$ lattices. One important aspect is that the lattice vectors can be described in terms of quaternions and the point group of the lattice has an interesting representation in terms of icosians. With this property it was noted that it is a sub-lattice in a quasicrystallographic lattice in 4D described by the point group symmetry $W(H_4)$. When projected orthogonally onto the Coxeter plane it has peculiar structures some of which have been shared with the quasicrystallographic structure obtained by the projection of the 5D cube with 32 vertices. Another interesting aspect is that the projected 5-cell has a one-to one correspondence with the de Brujin parameter $\zeta = \exp(i \frac{2\pi}{5})$ and its powers. Therefore the pentagrid structure introduced by de Brujin can be completely described by the projection method of the $A_4$ lattices. We have already pointed out that the projected Voronoi cell of the root lattice represents the nested decagram used as a framework for some of the Islamic arts. Perhaps most important observation is the existence of the point distributions on the Coxeter plane representing inflating decagram structures and the inflating rhombus structures.

We anticipate that the multigrid structure introduced by de Brujin can be reproduced by the projection technique of the lattices described by the affine Coxeter groups, in particular, these will be simpler with the $A_n$ series.

**Appendix A:** The list of 120 vectors created by the Coxeter-Weyl group
$W(A_4)(a_1\omega_1 + a_2\omega_2 + a_3\omega_3 + a_4\omega_4)$

| | | | |
|---|---|---|---|
| $a_2+a_3+a_4$ | $-a_4$ | $-a_1-a_2-a_3$ | $a_1+a_2$ |
| $a_1+a_2+a_3+a_4$ | $-a_3-a_4$ | $-a_2$ | $a_2+a_3$ |
| $-a_1-a_2$ | $a_1+a_2+a_3$ | $a_4$ | $-a_2-a_3-a_4$ |
| $-a_2-a_3$ | $a_2$ | $a_3+a_4$ | $-a_1-a_2-a_3-a_4$ |
| $-a_3$ | $-a_2$ | $-a_1$ | $a_1+a_2+a_3+a_4$ |
| $a_3$ | $-a_1-a_2-a_3$ | $a_1$ | $a_2+a_3+a_4$ |
| $a_2+a_3$ | $a_4$ | $-a_3-a_4$ | $-a_1-a_2$ |
| $-a_1-a_2-a_3-a_4$ | $a_1$ | $a_2$ | $a_3$ |
| $-a_2-a_3-a_4$ | $-a_1$ | $a_1+a_2+a_3$ | $-a_3$ |
| $-a_2$ | $a_2+a_3+a_4$ | $-a_1-a_2-a_3-a_4$ | $a_1+a_2+a_3$ |
| $a_1$ | $a_2+a_3$ | $-a_3$ | $a_3+a_4$ |
| $-a_1-a_2-a_3$ | $a_1+a_2+a_3+a_4$ | $-a_2-a_3-a_4$ | $a_2$ |
| $-a_3-a_4$ | $a_3$ | $-a_2-a_3$ | $-a_1$ |
| $-a_4$ | $-a_3$ | $-a_1-a_2$ | $a_1$ |
| $a_4$ | $-a_1-a_2-a_3-a_4$ | $a_1+a_2$ | $-a_2$ |
| $a_3+a_4$ | $-a_2-a_3-a_4$ | $a_2+a_3$ | $-a_1-a_2-a_3$ |
| $a_1+a_2+a_3$ | $-a_2-a_3$ | $a_2+a_3+a_4$ | $-a_3-a_4$ |
| $-a_1$ | $a_1+a_2$ | $a_3$ | $a_4$ |
| $a_2$ | $-a_1-a_2$ | $a_1+a_2+a_3+a_4$ | $-a_4$ |
| $-a_2$ | $a_2+a_3$ | $-a_1-a_2-a_3$ | $a_1+a_2+a_3+a_4$ |
| $a_1$ | $a_2+a_3+a_4$ | $-a_3-a_4$ | $a_3$ |
| $-a_1-a_2-a_3-a_4$ | $a_1+a_2+a_3$ | $-a_2-a_3$ | $a_2$ |
| $-a_3$ | $a_3+a_4$ | $-a_2-a_3-a_4$ | $-a_1$ |
| $a_4$ | $-a_2-a_3$ | $-a_1-a_2$ | $a_1$ |
| $-a_4$ | $-a_1-a_2-a_3$ | $a_1+a_2$ | $-a_2$ |
| $a_3$ | $-a_2-a_3$ | $a_2+a_3+a_4$ | $-a_1-a_2-a_3-a_4$ |
| $a_1+a_2+a_3+a_4$ | $-a_2-a_3-a_4$ | $a_2+a_3$ | $-a_3$ |
| $-a_1$ | $a_1+a_2$ | $a_3+a_4$ | $-a_4$ |
| $a_2$ | $-a_1-a_2$ | $a_1+a_2+a_3$ | $a_4$ |
| $-a_2-a_3$ | $a_2+a_3+a_4$ | $-a_1-a_2-a_3-a_4$ | $a_1+a_2$ |
| $a_1$ | $a_2$ | $a_3$ | $a_4$ |
| $-a_1-a_2$ | $a_1+a_2+a_3+a_4$ | $-a_2-a_3-a_4$ | $a_2+a_3$ |
| $-a_4$ | $-a_3$ | $-a_2$ | $-a_1$ |
| $-a_3-a_4$ | $a_3$ | $-a_1-a_2-a_3$ | $a_1$ |
| $a_3+a_4$ | $-a_1-a_2-a_3-a_4$ | $a_1+a_2+a_3$ | $-a_2-a_3$ |
| $a_4$ | $-a_2-a_3-a_4$ | $a_2$ | $-a_1-a_2$ |
| $a_1+a_2$ | $-a_2$ | $a_2+a_3+a_4$ | $-a_4$ |
| $-a_1$ | $a_1+a_2+a_3$ | $-a_3$ | $a_3+a_4$ |
| $a_2+a_3$ | $-a_1-a_2-a_3$ | $a_1+a_2+a_3+a_4$ | $-a_3-a_4$ |
| $-a_2-a_3$ | $a_2$ | $-a_1-a_2$ | $a_1+a_2+a_3+a_4$ |
| $a_1$ | $a_2+a_3+a_4$ | $-a_4$ | $-a_3$ |
| $-a_1-a_2-a_3-a_4$ | $a_1+a_2$ | $-a_2$ | $a_2+a_3$ |
| $a_3$ | $a_4$ | $-a_2-a_3-a_4$ | $-a_1$ |
| $a_3+a_4$ | $-a_4$ | $-a_1-a_2-a_3$ | $a_1$ |
| $-a_3-a_4$ | $-a_1-a_2$ | $a_1+a_2+a_3$ | $-a_2-a_3$ |
| $-a_3$ | $-a_2$ | $a_2+a_3+a_4$ | $-a_1-a_2-a_3-a_4$ |
| $a_1+a_2+a_3+a_4$ | $-a_2-a_3-a_4$ | $a_2$ | $a_1$ |
| $-a_1$ | $a_1+a_2+a_3$ | $a_4$ | $-a_3-a_4$ |
| $a_2+a_3$ | $-a_1-a_2-a_3$ | $a_1+a_2$ | $a_3+a_4$ |
| $-a_2-a_3-a_4$ | $a_2+a_3$ | $-a_1-a_2-a_3$ | $a_1+a_2$ |
| $a_1$ | $a_2$ | $a_3+a_4$ | $-a_4$ |
| $-a_1-a_2$ | $a_1+a_2+a_3$ | $-a_2-a_3$ | $a_2+a_3+a_4$ |
| $a_4$ | $-a_3-a_4$ | $-a_2$ | $-a_1$ |
| $-a_3$ | $a_3+a_4$ | $-a_1-a_2-a_3-a_4$ | $a_1$ |
| $a_3$ | $-a_1-a_2-a_3$ | $a_1+a_2+a_3+a_4$ | $-a_2-a_3-a_4$ |
| $-a_4$ | $-a_2-a_3$ | $a_2$ | $-a_1-a_2$ |
| $a_1+a_2$ | $-a_2$ | $a_2+a_3$ | $a_4$ |
| $-a_1$ | $a_1+a_2+a_3+a_4$ | $-a_3-a_4$ | $a_3$ |
| $a_2+a_3+a_4$ | $-a_1-a_2-a_3-a_4$ | $a_1+a_2+a_3$ | $-a_3$ |